\documentstyle[11pt,epsfig,glas]{article}
\begin{document}

\bibliographystyle{unsrt}    

\newcommand{\sst}{\scriptscriptstyle}
\newcommand{\mco}{\multicolumn}
\newcommand{\epp}{\epsilon^{\prime}}
\newcommand{\vep}{\varepsilon}
\newcommand{\ra}{\rightarrow}
\newcommand{\ppg}{\pi^+\pi^-\gamma}
\newcommand{\vp}{{\bf p}}
\newcommand{\ko}{K^0}
\newcommand{\kb}{\bar{K^0}}
\newcommand{\al}{\alpha}
\newcommand{\ab}{\bar{\alpha}}
\def\be{\begin{equation}}
\def\ee{\end{equation}}
\def\bea{\begin{eqnarray}}
\def\eea{\end{eqnarray}}
\def\CPbar{\hbox{{\rm CP}\hskip-1.80em{/}}}

\def\ap#1#2#3   {{\em Ann. Phys. (NY)} {\bf#1} (#2) #3.}
\def\apj#1#2#3  {{\em Astrophys. J.} {\bf#1} (#2) #3.}
\def\apjl#1#2#3 {{\em Astrophys. J. Lett.} {\bf#1} (#2) #3.}
\def\app#1#2#3  {{\em Acta. Phys. Pol.} {\bf#1} (#2) #3.}
\def\ar#1#2#3   {{\em Ann. Rev. Nucl. Part. Sci.} {\bf#1} (#2) #3.}
\def\cpc#1#2#3  {{\em Computer Phys. Comm.} {\bf#1} (#2) #3.}
\def\err#1#2#3  {{\it Erratum} {\bf#1} (#2) #3.}
\def\ib#1#2#3   {{\it ibid.} {\bf#1} (#2) #3.}
\def\jmp#1#2#3  {{\em J. Math. Phys.} {\bf#1} (#2) #3.}
\def\ijmp#1#2#3 {{\em Int. J. Mod. Phys.} {\bf#1} (#2) #3.}
\def\jetp#1#2#3 {{\em JETP Lett.} {\bf#1} (#2) #3.}
\def\jpg#1#2#3  {{\em J. Phys. G.} {\bf#1} (#2) #3.}
\def\mpl#1#2#3  {{\em Mod. Phys. Lett.} {\bf#1} (#2) #3.}
\def\nat#1#2#3  {{\em Nature (London)} {\bf#1} (#2) #3.}
\def\nc#1#2#3   {{\em Nuovo Cim.} {\bf#1} (#2) #3.}
\def\nim#1#2#3  {{\em Nucl. Instr. Meth.} {\bf#1} (#2) #3.}
\def\np#1#2#3   {{\em Nucl. Phys.} {\bf#1} (#2) #3.}
\def\pcps#1#2#3 {{\em Proc. Cam. Phil. Soc.} {\bf#1} (#2) #3.}
\def\pl#1#2#3   {{\em Phys. Lett.} {\bf#1} (#2) #3.}
\def\prep#1#2#3 {{\em Phys. Rep.} {\bf#1} (#2) #3.}
\def\prev#1#2#3 {{\em Phys. Rev.} {\bf#1} (#2) #3.}
\def\prl#1#2#3  {{\em Phys. Rev. Lett.} {\bf#1} (#2) #3.}
\def\prs#1#2#3  {{\em Proc. Roy. Soc.} {\bf#1} (#2) #3.}
\def\ptp#1#2#3  {{\em Prog. Th. Phys.} {\bf#1} (#2) #3.}
\def\ps#1#2#3   {{\em Physica Scripta} {\bf#1} (#2) #3.}
\def\rmp#1#2#3  {{\em Rev. Mod. Phys.} {\bf#1} (#2) #3.}
\def\rpp#1#2#3  {{\em Rep. Prog. Phys.} {\bf#1} (#2) #3.}
\def\sjnp#1#2#3 {{\em Sov. J. Nucl. Phys.} {\bf#1} (#2) #3.}
\def\spj#1#2#3  {{\em Sov. Phys. JEPT} {\bf#1} (#2) #3.}
\def\spu#1#2#3  {{\em Sov. Phys.-Usp.} {\bf#1} (#2) #3.}
\def\zp#1#2#3   {{\em Zeit. Phys.} {\bf#1} (#2) #3.}

\setcounter{secnumdepth}{2} 


\begin{titlepage}{GLAS--PPE/95--03}{\today}   
\title{Direct and resolved photoproduction at HERA \\ with virtual and
quasi-real photons}

\author{M.L. Utley} 
\centerline{talk given at the Europhysics Conference on High Energy Physics}
\centerline{27th July - 2nd August 1995}
\centerline{on behalf of the ZEUS collaboration}

\begin{abstract} 
Preliminary results are presented from a study of dijet photoproduction in ep
collisions with both virtual and quasi-real photons at the ZEUS detector.
Samples of events with photons of virtuality $P^2$ in the ranges $0.1 < P^2 <
0.55$ GeV$^2$ and $P^2 < 0.02$ GeV$^2$ having two jets of $E_T^{jet} > 4$ GeV in
the final state have been obtained.
For both quasi-real and virtual photons, uncorrected distributions of the
quantity $x_{\gamma}^{obs}$, the fraction of the photon momentum manifest in the
two highest $E_T$ jets, are presented. These distributions are sensitive to the
relative contributions of the direct and resolved processes. Resolved photon
processes are evident in both data sets, with an apparent decrease in the
relative contribution from resolved processes as photon virtuality increases.

\end{abstract}

\end{titlepage}

\section{Introduction}
\begin{figure}[hbt]
\begin{center}\mbox{\epsfig{file=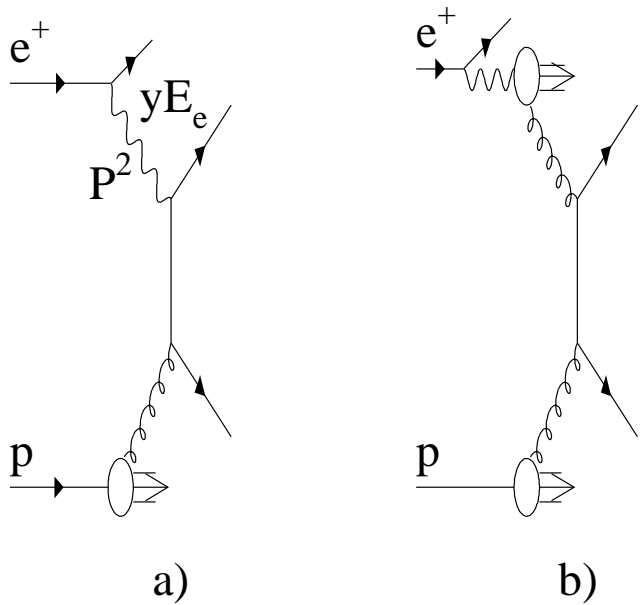,width= 12cm}}\end{center}
\caption{Diagrams showing a) direct and b) resolved photon processes. In both
cases the photon of virtuality $P^2$ carries a fraction $y$ of the positron
momentum.}
\label{fig:diresfig}
\end{figure}

Leading order (LO) QCD predicts photon interactions to have a two-component
nature. In direct photon processes the whole of the photon takes part in the
hard subprocess with a parton from the proton whereas in resolved photon
processes, the photon acts as a source of partons and one of these enters the
hard subprocess (see Figure \ref{fig:diresfig}). Both LO processes are
characterized by having two outgoing partons of large transverse energy.
Previous studies of dijet photoproduction at HERA have shown that both classes
of process are evident for the case of quasi-real photons (those of negligible
virtuality $P^2$) \cite{prevpap}.
 The parton content of photons is neither well constrained theoretically nor
well known experimentally, particularly for photons with small but non-zero
virtualities. Various theoretical predictions exist for the behaviour of the
photon structure as a function of the photon virtuality \cite{theory}. The
general expectation is that the contribution to the dijet cross section of
resolved photon processes should decrease relative to the contribution from
direct photon processes ({\it i.e.} that the partonic content of the photon is
suppressed) as the virtuality of the photon increases.

\section{The Data}
The data used in this analysis were collected during the 1994 run when HERA
collided 27.5 GeV positrons with 820 GeV protons. The ZEUS detector is described
elsewhere\cite{standard}. A tungsten-silicon sampling calorimeter was installed
for the 1994 running period. This 'beampipe' calorimeter tagged positrons
scattered through small angles (17 - 35 mrad) and gave a sample of events with
photon virtualities in the range $0.1 < P^{2} < 0.55 $ GeV$^2$. A sample of
events with quasi-real photons ($P^2 < 0.02$ GeV$^2$ with a median of $10^{-5}$
GeV$^2$) was obtained by requiring that the scattered positron be detected in
the downstream luminosity calorimeter\cite{lumi}. Jets were found in the main
uranium-scintillator\cite{main} calorimeter using a cone algorithm \cite{cone}
in $\eta$ - $\phi$ space, where ${\phi}$ is azimuth and the pseudo rapidity
${\eta} = -{\ln}(\tan({\theta}/2))$, ${\theta}$ being defined with respect to
the proton direction. Only those events with two or more jets of transverse
energy $E_T^{jet} > 4$ GeV in the range $-1.125 < \eta^{jet} <  1.875 $ were
selected as dijet events.
To reduce contamination from beam gas and deep inelastic scattering events, a
cut of $ 0.15 < y_{JB}=\sum(E-p_{z})/2E_{e} < 0.70 $ was applied, where the sum
is over calorimeter cells with deposits of total and longitudinal energy $E$ and
$p_z$ respectively and where $E_e$ is the positron beam energy. In
photoproduction, $y$ is the fraction of the positron energy carried by the
photon. These cuts left a sample of 375 events with virtual photons and a sample
of 14181 events with quasi-real photons, corresponding to respective integrated
luminosities of 2.07 pb$^{-1}$ and 2.19 pb$^{-1}$. 

\section{Results}

For each dijet event the fraction of the photon momentum manifest in the two
highest $E_T$ jets, $x_{\gamma}^{obs}$ was calculated. $x_{\gamma}^{obs}$ is
defined by

\[
x_{\gamma}^{obs} = \frac{ \sum_{jets}E_T^{jet}e^{-\eta^{jet}}}{2yE_e}
\]

\begin{figure}[hbt]
\begin{center}\mbox{\epsfig{file=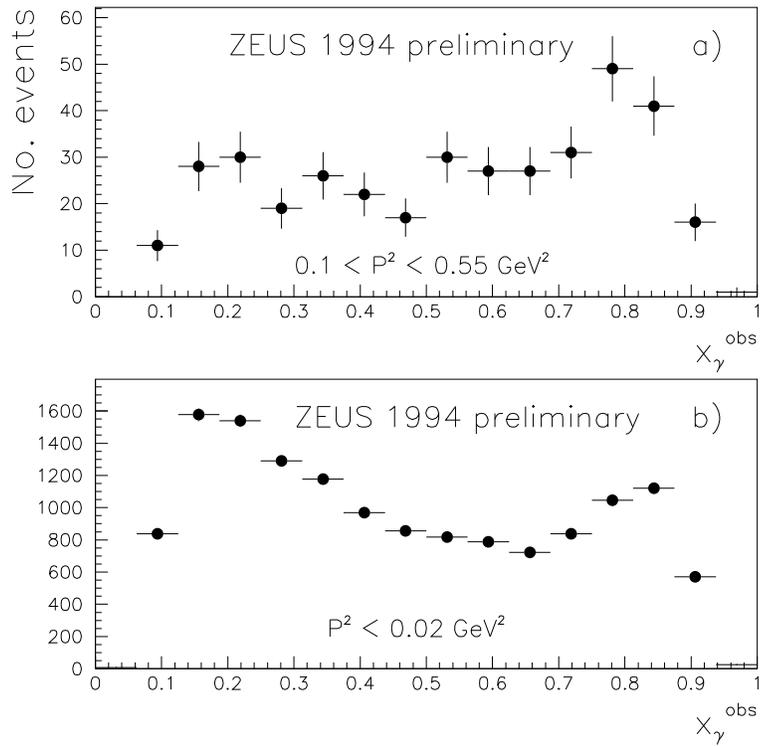,width= 11cm}}\end{center}
\caption{Uncorrected $x_{\gamma}^{obs}$ distributions for a) virtual and b)
quasi-real photons}
\label{fig:xgfig}
\end{figure}

\begin{figure}[hbt]
\begin{center}\mbox{\epsfig{file=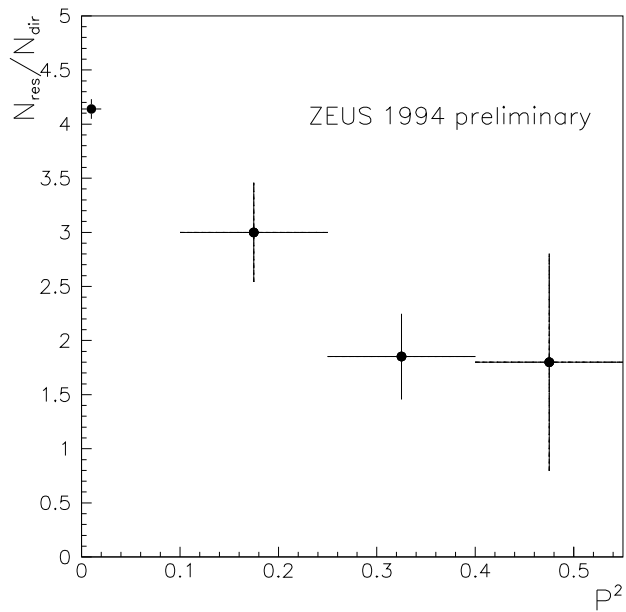,width= 9cm}}\end{center}
\caption{The uncorrected ratio $N_{res}/N_{dir}$ as a function of photon
virtuality $P^2$ (GeV$^2$)}
\label{fig:resdirfig}
\end{figure}

We measure $E_T^{jet}$ and ${\eta}^{jet}$ using the raw calorimeter energies and
use the Jaquet-Blondel method, $y_{JB}$ above, to measure $y$. No corrections
are made for detector effects.   
Uncorrected distributions of $x_{\gamma}^{obs}$ are shown in figure
\ref{fig:xgfig} a) for events with virtual photons and figure \ref{fig:xgfig} b)
for events with quasi-real photons. Events at high $x_{\gamma}^{obs}$ are
associated with direct photon processes while those at low $x_{\gamma}^{obs}$
are associated with resolved photon processes. It is clear that both classes of
event are present in both $P^2$ ranges. To quantify the relative contributions
of these classes of event, we have calculated the ratio $N_{res}/N_{dir}$,
defined as the number of events at low $x_{\gamma}^{obs}$ ($x_{\gamma}^{obs} <
0.75$) divided by the number of events at high $x_{\gamma}^{obs}$
($x_{\gamma}^{obs} > 0.75$). Figure \ref{fig:resdirfig} shows $N_{res}/N_{dir}$
as a function of $P^2$. $N_{res}/N_{dir}$ is independent of $y_{JB}$ for the
sample of events with quasi-real photons passing all the cuts applied. This
implies that the difference in the  $x_{\gamma}^{obs}$ distributions for the two
samples is not due to the differing $y_{JB}$ acceptances of the two positron
detectors.

The acceptance corrections that will eventually be applied to these data are
only weakly dependent on $x_{\gamma}^{obs}$ and we therefore expect the
corrections to $N_{res}/N_{dir}$ to be small. More work is required to
understand fully these corrections however. In conclusion we find the
preliminary result that for events with photons of virtuality $0.1 < P^2 < 0.55$
GeV$^2$ there is a contribution from resolved photon processes. The size of this
contribution relative to that from direct photon processes seems to decrease
with increasing photon virtuality.


\setcounter{secnumdepth}{0} 


\end{document}